\documentclass[runningheads,dvipsnames]{llncs}

\usepackage{amstext}
\usepackage[T1]{fontenc}
\usepackage{graphicx}
\usepackage[utf8]{inputenc}

\usepackage{subcaption}
\usepackage{amsmath}
\usepackage{amssymb}

\usepackage{xspace}
\usepackage{turnstile}

\usepackage{fontawesome5}

\usepackage{hyperref}
\usepackage{booktabs}
\usepackage{multirow}
\usepackage{pifont}

\usepackage{multicol}
\usepackage{stackengine}
\usepackage{pgf}

\usepackage{tikz}
\usetikzlibrary{decorations.pathreplacing,positioning}
\usetikzlibrary{arrows,automata}
\tikzset{>=stealth, shorten >=1pt}
\tikzset{every edge/.style = {thick, ->, draw}}
\tikzset{every loop/.style = {thick, ->, draw}}
\pgfdeclarelayer{background}
\pgfsetlayers{background,main}
\usetikzlibrary{petri,arrows,positioning}

\usetikzlibrary{
  calc,
  arrows,
  positioning,
  shapes,
  shadows,
  automata,
  external,
  shadows,
  decorations.pathreplacing,
  decorations.pathmorphing,
  decorations.markings,
  backgrounds,
  patterns
}

\tikzset{  
  node distance=4em,
  >=stealth',
  shadow/.style		= {opacity=.25, black, shadow xshift=0.08, shadow yshift=-0.08},
  plainnode/.style 	= {draw, ultra thick, fill=gray!10},
  pl0/.style			= {circle, minimum size=10mm, plainnode, drop shadow = {shadow}}
}

\usepackage[color=Cyan!50]{todonotes}

\newcommand{\nats}{\mathbb{N}}
\renewcommand{\epsilon}{\varepsilon}
\renewcommand{\phi}{\varphi}

\newcommand{\set}[1]{\{#1\}}

\newcommand{\F}{\mathop{\mathbf{F}\vphantom{a}}\nolimits}
\newcommand{\G}{\mathop{\mathbf{G}\vphantom{a}}\nolimits}
\DeclareMathOperator{\U}{\mathbf{U}}
\newcommand{\X}{\mathop{\mathbf{X}\vphantom{a}}\nolimits}

\newcommand{\ltl}{{LTL}\xspace}
\newcommand{\ctl}{{CTL}\xspace}

\newcommand{\ctlstar}{{CTL$^*$}\xspace}
\newcommand{\hyltl}{{HyperLTL}\xspace}
\newcommand{\hyctlstar}{{HyperCTL$^*$}\xspace}

\newcommand{\enabled}[1]{\texttt{en}_{#1}}

\begin{document}

\title{TAPAAL HyperLTL: A Tool for Checking Hyperproperties of Petri Nets}
%
%
\author{
Bruno Maria René Gonzalez\inst{1}\orcidID{0009-0006-8122-7160} \and
Peter Gjøl Jensen\inst{2}\orcidID{0000-0002-9320-9991} \and
Stefan Schmid\inst{1}\orcidID{0000-0002-7798-1711} \and
Ji\v{r}\'{\i} Srba\inst{2}\orcidID{0000-0001-5551-6547} \and
Martin Zimmermann\inst{2}\orcidID{0000-0002-8038-2453}
}
\authorrunning{Gonzalez et al.}
%
\institute{TU Berlin, Berlin, Germany
\and
Aalborg University, Aalborg, Denmark
\\
}
\maketitle              
\begin{abstract}
Petri nets are a modeling formalism capable of describing complex distributed systems and there exists a large number of both academic and industrial tools that enable automatic verification of model properties. Typical questions include reachability analysis and model checking against logics like LTL and CTL. However, these logics fall short when describing properties like non-interference and observational determinism that require simultaneous reasoning about multiple traces of the model and can thus only be expressed as hyperproperties. 
We introduce, to the best of our knowledge, the first HyperLTL model checker for Petri nets. The tool is fully integrated into the verification framework TAPAAL and we describe the semantics of the hyperlogic, present the tool's architecture and GUI, and evaluate the performance of the HyperLTL verification engine on two benchmarks of problems originating from the computer networking domain. 
\end{abstract}

\section{Introduction}

Many important properties of systems inherently relate multiple execution traces of a system, e.g., security and information-flow properties~\cite{if1,if2,if3,DBLP:conf/sas/TerauchiA05,if5,if6} as well as network properties like congestion~\cite{chiesa2016traffic,fortz2000internet}. 
These are not expressible in classical specification languages like \ltl~\cite{Pnueli77}, \ctl~\cite{DBLP:conf/lop/ClarkeE81}, and \ctlstar~\cite{DBLP:journals/jacm/EmersonH86}, as those are restricted to reasoning about one trace at a time. 
Clarkson and Schneider termed properties relating multiple traces \emph{hyperproperties} and initiated their rigorous investigation~\cite{DBLP:journals/jcs/ClarksonS10}.
Technically speaking, a hyperproperty is a set of sets of traces, just like a trace property is a set of traces. 
Their study received considerable attention after the introduction of specification languages for hyperproperties, which enabled the specification, analysis, and verification of hyperproperties.
The two most important ones are \hyltl and \hyctlstar which extend \ltl and \ctlstar by quantification over traces~\cite{DBLP:conf/post/ClarksonFKMRS14}. 
These logics are able to express many important hyperproperties from security like non-interference, non-inference, observational determinism, etc.~\cite{DBLP:conf/cav/FinkbeinerRS15}.
On the other hand, they are also able to express properties about paths in graphs, e.g., networks, like the existence of several disjoint paths between a source and a target node.
This allows us to formalize quantitative aspects like congestion using \hyltl
as a~requirement on the maximal number of flows that can traverse any given edge.

Petri nets~\cite{Petri:PhD} are widely used to represent concurrent and distributed systems due to their expressive power and 
an intuitive graphical representation.
Despite the versatility of Petri nets, no prior \hyltl verification tool has provided user-friendly support 
for designing and verifying Petri net models. Existing approaches often rely on textual specifications or lack intuitive interfaces. 

To address this challenge, we introduce TAPAAL \hyltl, a novel \hyltl model checker integrated into the TAPAAL~\cite{DJJJMS:TACAS:12} verification suite, specifically designed to verify complex temporal properties of distributed systems modeled as Petri nets. Our implementation is the first to bring \hyltl verification to Petri nets, offering a robust verification engine coupled with an intuitive user interface for modeling as well as debugging purposes.
                                                                    
To evaluate our tool, we conduct an extensive case study showing the applicability of \hyltl for the
analysis of congestion and latency in a computer networking setting.
Our results show that our \hyltl engine outperforms the baseline approach based on self-composition~\cite{DBLP:journals/mscs/BartheDR11,DBLP:conf/sas/TerauchiA05} and 
achieves competitive performance compared to state-of-the-art tools like MCHyper~\cite{DBLP:conf/cav/FinkbeinerRS15}.

This paper extends a tool paper presented at ATVA 2025~\cite{GJSSZ:ATVA:2025} with an additional case study dealing with latency in computer networks.

\paragraph{Related work.}
\hyltl and its branching-time companion \hyctlstar have been introduced and their model-checking problems have been shown decidable in the seminal work of Clarkson et al.~\cite{DBLP:conf/post/ClarksonFKMRS14}. 
In general, model checking of \hyltl is TOWER-complete~\cite{DBLP:phd/dnb/Rabe16,DBLP:conf/csl/Mascle020} in the number of quantifier alternations. 
Hence, almost all tool development has been concerned with the alternation-free fragment, although recently the first tools tackling (a small number of) alternations have been presented. 

For example, the tool MCHyper models the system using And-Inverter Graphs (AIGs) and has originally been restricted to alternation-free formulae~\cite{DBLP:conf/cav/FinkbeinerRS15} (like our tool), where it relies on the ABC~\cite{abc} backend.
More recently, it has been extended to handle one alternation using a game-based approach~\cite{DBLP:conf/cav/CoenenFST19}.
On the other hand, the tool AutoHyper handles quantifier alternations~\cite{DBLP:conf/tacas/BeutnerF23} by implementing an automata-theoretic model checking algorithm, relying on efficient automata inclusion checking.

Another approach for handling the inherent complexity of \hyltl model checking is to consider incomplete methods like bounded model checking, which searches for counterexamples of bounded size. 
Hsu et al.~\cite{DBLP:conf/tacas/HsuSB21,DBLP:conf/tacas/HsuBFS23} implemented this in their tool HyperQube using a reduction to QBF.

Finally, model checking asynchronous extensions of \hyltl has been studied by Baumeister et al.~\cite{DBLP:conf/cav/BaumeisterCBFS21} and probabilistic extensions by Dode et al.~\cite{DBLP:conf/fm/DobeABB21}.
Most recently, the game-based approach mentioned above has been generalized~\cite{BF,prophy} and planning-based~\cite{planning} algorithms and implementations have been presented.

None of the existing tools mentioned above can handle Petri nets natively. 
Thus, our tool offers an alternative modeling language based on Petri nets, which naturally support concurrency, while existing tools use NuSMV~\cite{DBLP:conf/cav/CimattiCGGPRST02} models (like HyperQube) or VHDL~\cite{vhdl} and VeriLog~\cite{verilog} models (like MCHyper).

\section{Modeling Formalism and \hyltl Logic}
\label{sec:prels}

We shall now semi-formally introduce the Petri net model as well as the syntax and semantics of the variant of \hyltl that is supported by our tool and tailored to express properties of Petri nets.

\subsection{Petri nets}

Our tool uses the classical Petri net (PN) model~\cite{Petri:PhD} with weighted and inhibitor arcs. It also supports
colored Petri nets, following the PNML syntax used in the annual Model Checking Contest (MCC)~\cite{mcc:2025}.
The colored PNs are unfolded into classical P/T (place/transition) nets, after which the \hyltl model checking is executed.

\begin{figure}[t!]
\begin{subfigure}[b]{0.25\textwidth}
    \centering
     \scalebox{0.8}{
    \begin{tikzpicture}[thick, xscale=0.9, yscale=1.8]
    \draw[white] (-1,-3.3) rectangle (1,0.3);
    \node[label=above:$v_0$] (v0) at (0,0) {\Large\faServer};
    \node[label=below:$v_1$] (v1) at (0,-2) {\Large\faServer};
    \node[label=left:$v_2$] (v2) at (-1.5,-1) {\Large\faServer};
    \node[label=left:$v_3$] (v3) at (1.5,-1) {\Large\faServer};

    \draw[-stealth]
     (v0) edge[bend left=10] (v1)
     (v0) edge[bend left=10] (v2)
     (v0) edge[bend left=10] (v3)
     (v1) edge[bend left=10] (v3)
     (v1) edge[bend left=10] (v2)
     (v1) edge[bend left=10] (v0)
     (v2) edge[bend left=10] (v0)
     (v3) edge[bend left=10] (v0)
     (v3) edge[bend left=10] (v1)
     (v2) edge[bend left=10] (v1);
    
    \end{tikzpicture} 
    }
   \caption{A comp. network}
    \label{fig:network}
\end{subfigure} 
\begin{subfigure}[b]{0.75\textwidth}
         \centering
         \scalebox{0.78}{
\centering
\begin{tikzpicture}[font=\footnotesize, xscale=0.29, yscale=0.37, x=1.33pt, y=1.33pt]
\tikzstyle{arc}=[->,>=stealth,thick]
\tikzstyle{transportArc}=[->,>=diamond,thick]
\tikzstyle{inhibArc}=[->,>=o,thick]
\tikzstyle{every place}=[minimum size=6mm,thick]
\tikzstyle{every transition} = [fill=black,minimum width=2mm,minimum height=5mm]
\tikzstyle{every token}=[fill=white,text=black]
\tikzstyle{sharedplace}=[place,minimum size=7.5mm,dashed,thin]
\tikzstyle{sharedtransition}=[transition, fill opacity=0, minimum width=3.5mm, minimum height=6.5mm,dashed]
\tikzstyle{urgenttransition}=[place,fill=white,minimum size=2.0mm,thin]
\tikzstyle{uncontrollabletransition}=[transition,fill=white,draw=black,very thick]
\tikzstyle{globalBox} = [draw,thick,align=left]
\node[place, label={[align=left,label distance=0cm]90:$v_0$}] at (480,-165) (V0) {};
\node at (480.0,-165.0)[circle,fill,inner sep=1.0pt]{};
\node[place, label={[align=left,label distance=0cm]355:${v_1}$}] at (480,-615) (V1) {};
\node[place, label={[align=left,label distance=0cm]90:${v_2}$}] at (128,-383) (V2) {};
\node[place, label={[align=left,label distance=0cm]90:${v_3}$}] at (833,-390) (V3) {};
\node[place, label={[align=left,label distance=0cm]270:${a({t_0})}$}] at (563,-390) (aT0) {};
\node at (563.0,-390.0)[circle,fill,inner sep=1.0pt]{};
\node[place, label={[align=left,label distance=0cm]270:$a(t'_0)$}] at (398,-390) (aT0_R) {};
\node at (398.0,-390.0)[circle,fill,inner sep=1.0pt]{};
\node[place, label={[align=left,label distance=0cm]90:${a(t_1)}$}] at (285,-180) (aT1) {};
\node at (285.0,-180.0)[circle,fill,inner sep=1.0pt]{};
\node[place, label={[align=left,label distance=0cm]180:${a(t'_1)}$}] at (308,-368) (aT1_R) {};
\node at (308.0,-368.0)[circle,fill,inner sep=1.0pt]{};
\node[place, label={[align=left,label distance=0cm]90:$a(t_2)$}] at (675,-180) (aT2) {};
\node at (675.0,-180.0)[circle,fill,inner sep=1.0pt]{};
\node[place, label={[align=left,label distance=0cm]0:${a(t'_2)}$}] at (638,-360) (aT2_R) {};
\node at (638.0,-360.0)[circle,fill,inner sep=1.0pt]{};
\node[place, label={[align=left,label distance=0cm]180:${a(t_3)}$}] at (308,-420) (aT3) {};
\node at (308.0,-420.0)[circle,fill,inner sep=1.0pt]{};
\node[place, label={[align=left,label distance=0cm]270:$a(t'_3)$}] at (263,-585) (aT3_R) {};
\node at (263.0,-585.0)[circle,fill,inner sep=1.0pt]{};
\node[place, label={[align=left,label distance=0cm]270:$a(t'_4)$}] at (683,-593) (aT4_R) {};
\node at (683.0,-593.0)[circle,fill,inner sep=1.0pt]{};
\node[place, label={[align=left,label distance=0cm]0:$a(t_4)$}] at (638,-420) (aT4) {};
\node at (638.0,-420.0)[circle,fill,inner sep=1.0pt]{};
\node[place, label={[align=left,label distance=0cm]180:$v'_1$}] at (420,-675) (V1_reached) {};
\node[place, label={[align=left,label distance=0cm]270:$
v'_2$}] at (128,-503) (V2_reached) {};
\node[transition, label={[align=left,label distance=0cm]120:$t_0$}] at (518,-390) (T0) {};
\node[transition, label={[align=left,label distance=0cm]330:$t'_0$}] at (443,-390) (T0_R) {};
\node[transition, label={[align=left,label distance=0cm]170:$t_1$}] at (285,-240) (T1) {};
\node[transition, label={[align=left,label distance=0cm]10:$t_2$}] at (675,-240) (T2) {};
\node[transition, label={[align=left,label distance=0cm]90:$t'_2$}] at (638,-300) (T2_R) {};
\node[transition, label={[align=left,label distance=0cm]10:$t_3$}] at (308,-480) (T3) {};
\node[transition, label={[align=left,label distance=0cm]170:$t_4$}] at (638,-480) (T4) {};
\node[transition, label={[align=left,label distance=0cm]90:$t'_4$}] at (683,-533) (T4_R) {};
\node[transition, label={[align=left,label distance=0cm]180:$t'_3$}] at (263,-525) (T3_R) {};
\node[transition, label={[align=left,label distance=0cm]90:$t'_1$}] at (308,-308) (T1_R) {};
\node[transition, label={[align=left,label distance=0cm]0:$v_1^{\mathit{deliver}}$}] at (480,-675) (V1_deliver) {};
\node[transition, label={[align=left,label distance=0cm]180:$v_2^{\mathit{deliver}}$}] at (128,-443) (V2_deliver) {};
\draw[arc,pos=0.5] (V0) to node[bend right=0,auto,align=left] {} (T0);
\draw[arc,pos=0.5] (V0) to node[bend right=0,auto,align=left] {} (T1);
\draw[arc,pos=0.5] (V0) to node[bend right=0,auto,align=left] {} (T2);
\draw[arc,pos=0.5] (V1) to node[bend right=0,auto,align=left] {} (T0_R);
\draw[arc,pos=0.5] (V1) to node[bend right=0,auto,align=left] {} (T3);
\draw[arc,pos=0.5] (V3) to node[bend right=0,auto,align=left] {} (T2_R);
\draw[arc,pos=0.5] (aT0) to node[bend right=0,auto,align=left] {} (T0);
\draw[arc,pos=0.5] (aT0_R) to node[bend right=0,auto,align=left] {} (T0_R);
\draw[arc,pos=0.5] (aT1) to node[bend right=0,auto,align=left] {} (T1);
\draw[arc,pos=0.5] (aT2) to node[bend right=0,auto,align=left] {} (T2);
\draw[arc,pos=0.5] (aT2_R) to node[bend right=0,auto,align=left] {} (T2_R);
\draw[arc,pos=0.5] (aT3) to node[bend right=0,auto,align=left] {} (T3);
\draw[arc,pos=0.5] (T0) to node[bend right=0,auto,align=left] {} (V1);
\draw[arc,pos=0.5] (T0_R) to node[bend right=0,auto,align=left] {} (V0);
\draw[arc,pos=0.5] (T1) to node[bend right=0,auto,align=left] {} (V2);
\draw[arc,pos=0.5] (T2) to node[bend right=0,auto,align=left] {} (V3);
\draw[arc,pos=0.5] (T2_R) to node[bend right=0,auto,align=left] {} (V0);
\draw[arc,pos=0.5] (T3) to node[bend right=0,auto,align=left] {} (V2);
\draw[arc,pos=0.5] (V3) to node[bend right=0,auto,align=left] {} (T4);
\draw[arc,pos=0.5] (T4) to node[bend right=0,auto,align=left] {} (V1);
\draw[arc,pos=0.5] (V1) to node[bend right=0,auto,align=left] {} (T4_R);
\draw[arc,pos=0.5] (T4_R) to node[bend right=0,auto,align=left] {} (V3);
\draw[arc,pos=0.5] (aT4) to node[bend right=0,auto,align=left] {} (T4);
\draw[arc,pos=0.5] (aT4_R) to node[bend right=0,auto,align=left] {} (T4_R);
\draw[arc,pos=0.5] (T3_R) to node[bend right=0,auto,align=left] {} (V1);
\draw[arc,pos=0.5] (aT3_R) to node[bend right=0,auto,align=left] {} (T3_R);
\draw[arc,pos=0.5] (V2) to node[bend right=0,auto,align=left] {} (T3_R);
\draw[arc,pos=0.5] (T1_R) to node[bend right=0,auto,align=left] {} (V0);
\draw[arc,pos=0.5] (aT1_R) to node[bend right=0,auto,align=left] {} (T1_R);
\draw[arc,pos=0.5] (V2) to node[bend right=0,auto,align=left] {} (T1_R);
\draw[arc,pos=0.5] (V1_deliver) to node[bend right=0,auto,align=left] {} (V1_reached);
\draw[arc,pos=0.5] (V1) to node[bend right=0,auto,align=left] {} (V1_deliver);
\draw[arc,pos=0.5] (V2) to node[bend right=0,auto,align=left] {} (V2_deliver);
\draw[arc,pos=0.5] (V2_deliver) to node[bend right=0,auto,align=left] {} (V2_reached);

\end{tikzpicture}
        }
    \caption{A Petri net $N$ modeling the computer network} 
    \label{fig:PN}
     \end{subfigure} \\[6mm]
   \begin{subfigure}[b]{1\textwidth}
$\varphi_1 \equiv \exists \pi_1.\ \exists \pi_2.\ (\F \pi_1.v'_1=1)  \wedge  (\F \pi_2.v'_1=1) \wedge \G \mathit{noCongestion2}$ \\[1mm]
$\varphi_2 \equiv \exists \pi_1.\ \exists \pi_2.\ \exists \pi_3.\ (\F \pi_1.v'_1=1)  \wedge  (\F \pi_2.v'_1=1) \wedge  (\F \pi_3.v'_1=1) \wedge \G \mathit{noCongestion3}$\\[1mm]
$\varphi_3 \equiv \exists \pi_1.\ \exists \pi_2.\ (\F \pi_1.v'_2=1)  \wedge  (\F \pi_2.v'_2=1) \wedge \G \mathit{noCongestion2}$\\[1mm]
$\varphi_4 \equiv \exists \pi_1.\ \exists \pi_2.\ \exists \pi_3.\ (\F \pi_1.v'_2=1)  \wedge  (\F \pi_2.v'_2=1) \wedge  (\F \pi_3.v'_2=1) \wedge \G \mathit{noCongestion3}$ \\[-2mm]

where \\[-2mm]

$\mathit{noCongestion2} \equiv \bigwedge_{t \in T \smallsetminus \{ v_1^\mathit{deliver}, v_2^\mathit{deliver}\}} (\pi_1.a(t) + \pi_2.a(t) \ge 1)$ 

$\mathit{noCongestion3} \equiv \bigwedge_{t \in T \smallsetminus \{ v_1^\mathit{deliver}, v_2^\mathit{deliver}\}} (\pi_1.a(t) + \pi_2.a(t) + \pi_3.a(t) \ge 2)$ 
\caption{Examples of \hyltl formulae where
$N \models \varphi_1$, $N \models \varphi_2$, $N \models \varphi_3$ and $N \not\models \varphi_4$}
\label{fig:formulae}
    \end{subfigure}
\caption{Example of a Petri net and \hyltl formulae}
\label{fig:example}
\end{figure}

Figure~\ref{fig:PN} shows an example of a P/T net where places from the set $P=\{v_0,\ldots,v_3,a(t_0),a(t'_0), \ldots, a(t_4), a(t'_4)\}$ are drawn as circles, transitions from the set $T=\{t_1, t'_1, \ldots, t_4, t'_4, v_1^\mathit{deliver}, v_2^\mathit{deliver}\}$ are drawn as rectangles, and arcs are the directed edges connecting either places to transitions or transitions to places. Unless otherwise stated, the default weight of all arcs is $1$.

A \emph{marking} $M\colon P \rightarrow \nats^0$ is a function that represents 
the placement of tokens (denoted as dots) across the places in the net.
A transition $t$ is \emph{enabled} in a~marking $M$ if there are enough tokens in all of the input places to the transition. An enabled transition~$t$ can \emph{fire} and produce the new marking $M'$,
written as $M [t\rangle M'$, by (i) removing
as many tokens from the input places as is the weight of the corresponding arc, and (ii) producing new tokens to every output place of the transition, again according to the weights of the output arcs.
For example, firing the transition~$t_1$ in Figure~\ref{fig:PN} removes the tokens from $v_0$ and $a(t_1)$ and adds a token to $v_2$. All other tokens are unchanged.

The Petri net in Figure~\ref{fig:PN} models all possible routing sequences for the computer network depicted
in Figure~\ref{fig:network} where a packet (token) starts at the node $v_0$ and the aim
is to reach the node $v_1$ or $v_2$. Moreover, every link in the network corresponds to
some transition $t$ in the Petri net and once this transition fires, the token in 
the place $a(t)$ is consumed, representing the fact that the corresponding link is now occupied.

A \emph{trace} (run) in a Petri net is an infinite sequence $\rho = M_0 M_1 M_2 \cdots$ of markings
such that for every $n \geq 0$ either (i) $M_n [t_n\rangle M_{n+1}$ for some transition
$t_n \in T$, or (ii) $M_{n+1} = M_n$ in case that $M_n$ is a deadlock, i.e., if $M_n$ does not enable
any transition. As \hyltl is interpreted over infinite traces, we introduce the stuttering 
to prolong possibly deadlocked traces into infinite ones (as it is e.g., assumed in the MCC~\cite{mcc:2025}).
Given a trace $\rho = M_0 M_1 M_2 \cdots$, we denote the $n$'th marking $M_n$ in the trace by $\rho^n$.

\subsection{HyperLTL}
\label{subsec_hyperltl}

\hyltl~\cite{DBLP:conf/post/ClarksonFKMRS14} extends \ltl~\cite{Pnueli77} (which is evaluated over single traces) by  quantification over multiple traces, and is therefore evaluated over sets of traces.
Our tool supports an alternation-free hyperlogic specifically tailored to Petri nets. 

Figure~\ref{fig:formulae} shows example formulae in the logic where the $\pi_i$ are trace variables ranging over infinite traces of a Petri net. 
Every formula starts with either existential or universal quantification
over a list of trace variables $\pi_1, \ldots, \pi_m$.
Thus, quantification assigns traces of the Petri net to the trace variables.
This is followed by a formula
composed of the standard \ltl temporal operators that is (synchronously) evaluated over the quantified traces:
\begin{itemize}
    \item $\X\varphi$ stating that $\varphi$ holds at the next position, 
    \item $\F\varphi$ stating that $\varphi$ holds at some future position,
    \item $\G\varphi$ stating that $\varphi$ holds at all future positions, and
    \item $\psi \U \varphi$ stating that $\varphi$ holds at some future position and $\psi$ holds at all intermediate positions.
\end{itemize}
Finally, we consider two types of atomic propositions:
\begin{itemize}
    \item for every variable~$\pi$ and every transition~$t$ there is a proposition~$\pi.\enabled{t}$, and
    \item we allow linear (in)equalities of the form~$\sum_{\ell} c_\ell \cdot \pi_{i_\ell}.p_\ell \bowtie b$ where the
    $c_\ell$ and $b$ are integer constants, the $\pi_{i_\ell}$ are trace variables, and the $p_\ell$ are places of the net, and where $\bowtie\, \in \set{<, \le , = , \ge, >}$ is a comparison operator.
\end{itemize}
Now, assuming that the trace~$\rho_i$ is assigned to the variable~$\pi_i$ for each $i$, we
evaluate atomic propositions at position~$n$ as follows:
\begin{itemize}
    \item $\pi_i.\enabled{t}$ is satisfied if $t$ is enabled in the $n$'th marking~$\rho^n_i$, and 
    \item $\sum_{\ell} c_\ell \cdot \pi_{i_\ell}.p_\ell \bowtie b$ is satisfied if the inequality obtained by replacing each $\pi_{i_\ell}.p_\ell$ with the number of tokens in the marking 
    $\rho_{i_\ell}^n$, i.e., the value~$\rho_{i_\ell}^n(p_\ell)$, is valid.
\end{itemize}
For a formal definition of the syntax and semantics of \hyltl, see, e.g.,~\cite{DBLP:conf/post/ClarksonFKMRS14}.

Coming back to our example in Figure~\ref{fig:example}, the formula~$\varphi_1$ (resp.\ $\varphi_3$) from Figure~\ref{fig:formulae} expresses that there are two traces that both eventually (but possibly at different positions) receive a token in $v_1'$  (resp. $v_2'$) and each transition $t$ is fired in at most one of the traces (hence there must
be a token in $a(t)$ in at least one of the two traces).
In other words, $\varphi_1$ and $\varphi_3$ express that there are two disjoint paths in the graph in Figure~\ref{fig:network}, starting at $v_0$ and leading to $v_1$ resp.\ $v_2$.
Analogously, $\varphi_2$ and $\varphi_4$  express similar properties, but requiring the existence
of three disjoint paths. 
Hence, $\varphi_1$, $\varphi_2$, and $\varphi_3$ are satisfied by the example net~$N$, but not $\varphi_4$.

The reason for introducing the places~$v'_1$ and $v'_2$ is that once the token initially in $v_0$ arrives to one of them, the corresponding trace gets stuck and the last reached marking is allowed to stutter and  hence does not use any of the remaining link capacities. This is important as the existentially quantified traces may be of different lengths before reaching the goal place, and such traces must globally synchronize.

\begin{figure}[t]
    \centering
    \includegraphics[width=0.95\linewidth]{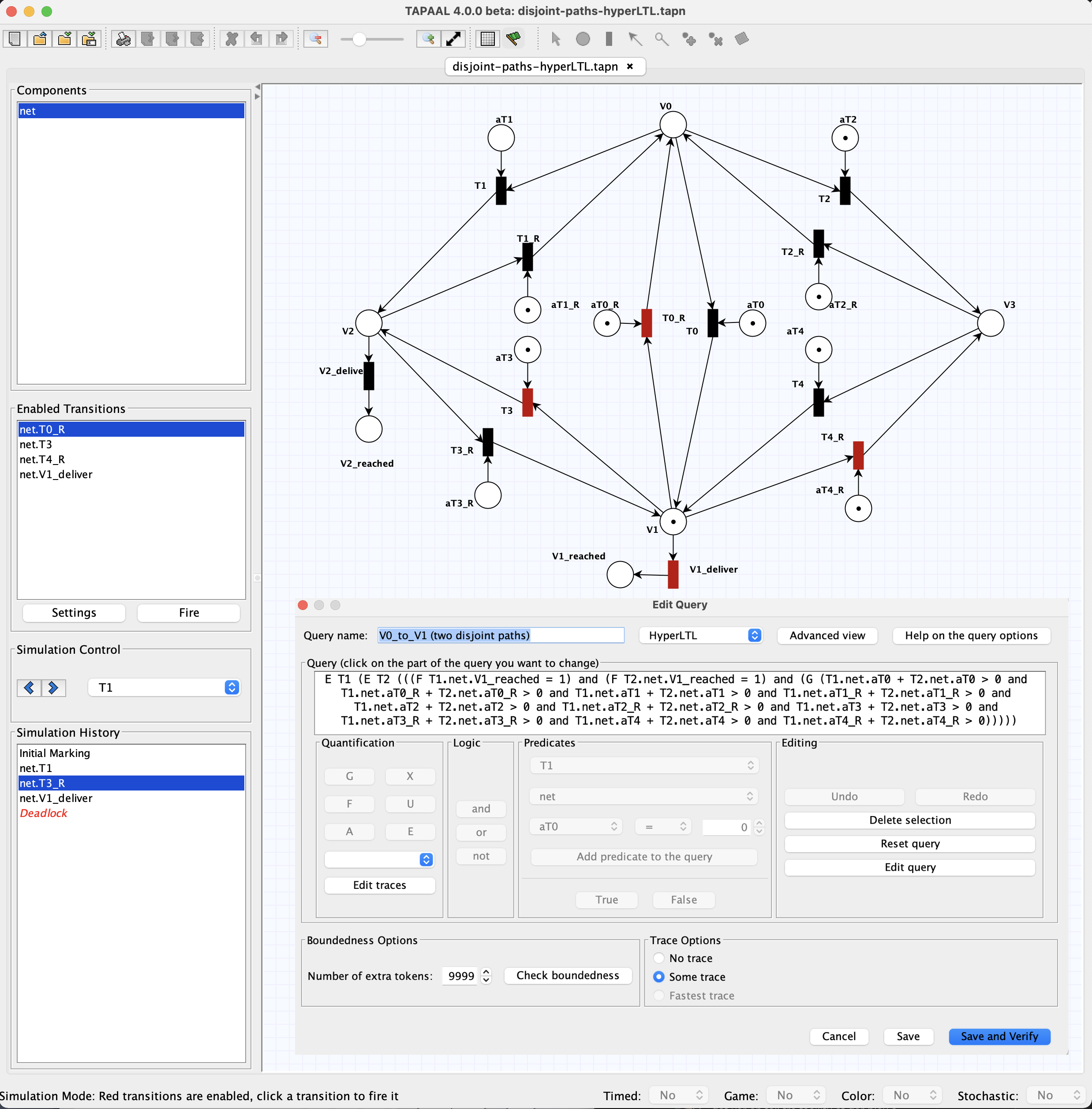}
    \caption{TAPAAL \hyltl screenshot (simulator mode with a query dialog)}
    \label{fig:screenshot}
\end{figure}


\section{Tool Implementation and Graphical User Interface}

The verification engine of our tool is implemented in C++ and extends
the existing \ltl verification engine that is part of the {\tt verifypn}
command line tool~\cite{JNOS:ToPNoC:16}. The \hyltl engine supports nets described in PNML~\cite{pnml} and, for universal formulae~$\varphi$, constructs
in an on-the-fly manner the vector of markings currently reached in all 
the considered traces and explores its product with the B\"uchi automaton
representing the negation of the \ltl property obtained by dropping the quantifiers of $\varphi$. On this product B\"uchi automaton,
we perform a classical search for a reachable accepting loop using the nested
DFS search strategy~\cite{ndfs}. If such a loop is found, the engine returns
the verification answer together with the set of traces (in an XML format)
that form such a loop. In the case where no counter-example exists, the tool reports that the property is 
satisfied together with statistics about the explored state-space.
Existential quantification is handled by negating the formula and swapping the results the tool reports.

The \hyltl engine is directly called from the tool TAPAAL~\cite{DJJJMS:TACAS:12} that 
has been extended with a graphical way to construct \hyltl queries as well as a simulator that allows to replay multiple traces returned by the engine.
Figure~\ref{fig:screenshot} displays a screenshot of the TAPAAL \hyltl interface.
The GUI is in simulation mode where the user can select the traces returned by the verification engine (currently, trace T1 is selected) and simulate the traces
in the GUI. A graphical dialog for creating \hyltl queries is shown as overlay.
The tool is available at \url{http://www.tapaal.net/downloads}, including
a complete reproducibility package~\cite{Repropackage}.


\begin{figure}[t]
    \centering
    \include{disjoint-paths-hyperLTL-selfcomposed-2}
    \caption{Self-composition with the \ltl query $(\F v_1^1=1) \wedge (\F v_1^2=1$)}
    \label{fig:PNself}
\end{figure}

\begin{figure}[t]
    \centering
    \includegraphics[width=0.8\linewidth]{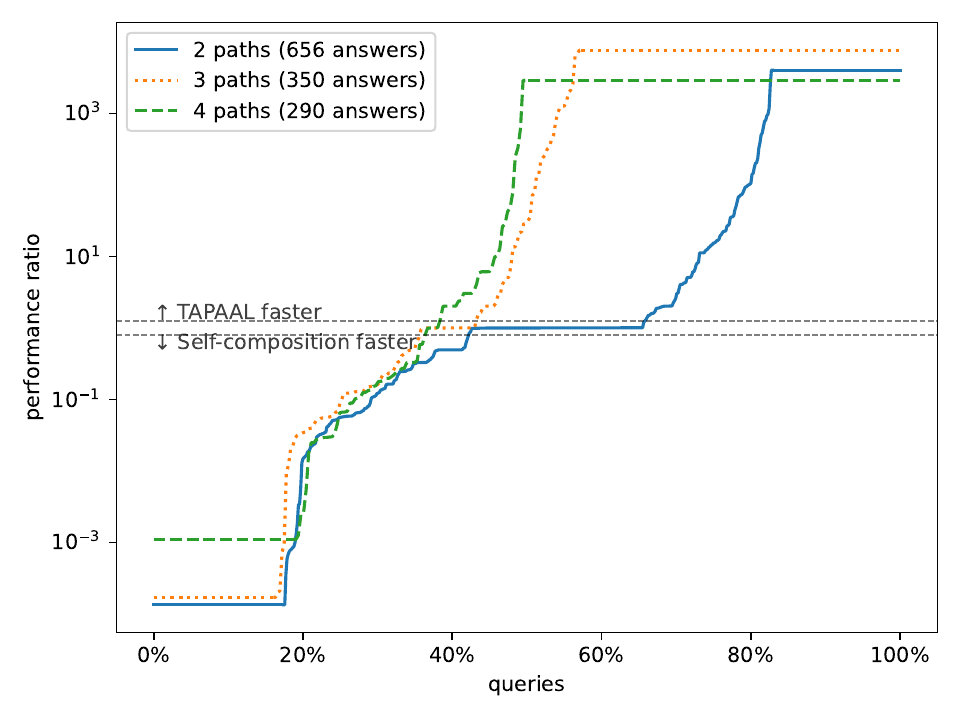}
    \caption{Ratio plot of TAPAAL \hyltl vs.\ Self-composition}
    \label{fig:tapaal-vs-selfcomposition}
\end{figure}

\section{Congestion Case Study}

\begin{figure}[t]
    \centering
    \includegraphics[width=0.8\linewidth]{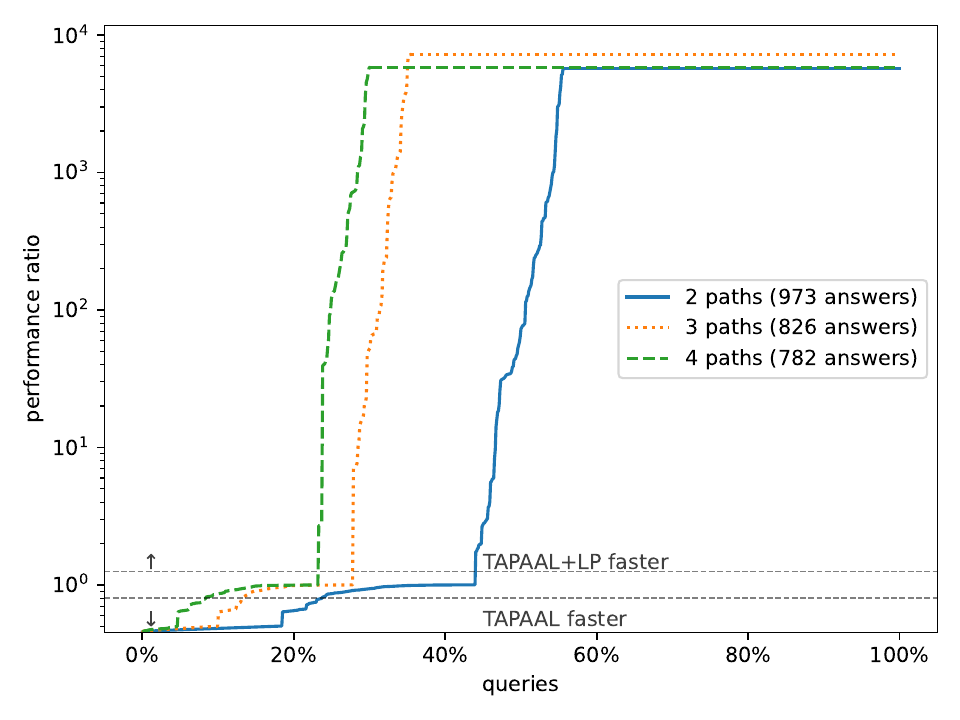}
    \caption{Ratio plot of TAPAAL \hyltl with and without LP Check}
    \label{fig:tapaal-vs-tapaal-LP}
\end{figure}

We evaluate the performance of our \hyltl model checker on a case study inspired by  routing problems from computer networking~\cite{dunn1994comparison}. For given source and destination nodes $s$ and $t$, and for a given network (directed graph), we want to find $k$ directed paths from $s$ to $t$ such that
these paths do not cause congestion on any of the links (edges) in the network.
In our simplified scenario, we say that an edge is congested if there are strictly more than
$\ell$ paths from $s$ to $t$ that use the given edge; hence $\ell$ indirectly models edge
capacities. The introductory example in Figure~\ref{fig:PN} shows how this problem can
be modeled as a Petri net. The formulae for $k=2,3$ and $\ell=1$ are depicted
in Figure~\ref{fig:formulae} as $\varphi_1$ and $\varphi_3$ for the target node $v_1$
and as $\varphi_2$ and $\varphi_4$ for the target node $v_2$.

Our benchmark contains 3900 \hyltl formulae, evaluated
on Petri net models of 260 real-world network topologies from the Topology Zoo dataset~\cite{zoo}. We consider three $(k,\ell)$ problem variants for
$(2,1)$, $(3,1)$ and $(4,2)$, where for each network topology we generate five \hyltl queries for randomly selected pairs of source and target nodes. To balance the number of true and false queries in the benchmark, the source is selected to be a random high-degree node.
The experiments are executed on an AMD EPYC 7551 processor running at 1996~MHz, with 900 seconds timeout and 2GB memory limit.

First, we compare our \hyltl implementation (referred to as TAPAAL) with
a self-composition approach~\cite{DBLP:journals/mscs/BartheDR11,DBLP:conf/sas/TerauchiA05} 
that creates a copy of the composed model for each trace and adds a synchronization
mechanism to the~model in order to guarantee that we iteratively perform one step in each copy of the~model before we evaluate the predicates and continue with another single
step in each copy. This allows us to reduce the \hyltl
formula into a normal LTL formula where instead of each trace we now refer to the
respective copy of the model. However, this is at the expense of creating a possibly
complicated model that explodes with the number of traces and additionally implements
a~synchronization mechanism in order to keep all copies synchronized.

In our concrete example, the self-composition does not require such a complicated
synchronization mechanism as for each quantified trace we can create
a copy of the net that can run completely concurrently (avoding the lock-step synchronization), while checking for
the congestion using the shared places $a(t)$ that contain as many tokens as
is the edge capacity. Figure~\ref{fig:PNself} shows our simplified self-composition for our running example as well as a classical \ltl query that expresses the same property as the \hyltl formula $\varphi_1$ from Figure~\ref{fig:formulae}. To verify the classical LTL formula
on the self-composed system, we benchmark our tool against the LTL engine of TAPAAL~\cite{JSUV:VMCAI:2022}, the winner 
in the 2025 Model Checking Contest~\cite{mcc:2025} in the LTL category.
A ratio plot is depicted in Figure~\ref{fig:tapaal-vs-selfcomposition} where the $x$-axis contains all queries solved by at least one of the methods, sorted by the ratio of self-composition running time
divided by TAPAAL \hyltl running time. 
We remark that because the plot contains three different problem instances where
the tools solve different number of queries (the numbers of answers in the
parenthesis show the total number of solved problems by at least one of the methods),
we use the percentage scale on the x-axis instead of the absolute count.
For two disjoint paths, both methods are comparable, however, for $3$ and $4$ paths there is a clear advantage of using our new \hyltl implementation. For example, for $4$ paths, the self-composition timeouts (depicted by the straight horizontal line) on more than 50\% of all queries that the \hyltl implementation managed to solve. This is in particular true for queries with positive answers, as the on-the-fly method that we implemented in TAPAAL \hyltl is more efficient than self-composition, where the net
size explodes with number of trace variables in the \hyltl formula.

In order to further improve the performance of our tool on negative queries, we employ
an over-approximation method based on state-equations and linear programming~\cite{BDJJS:PN:18} (we refer
to this method as TAPAAL+LP). We create the~self-composition net and apply the fast
LP check that can in many cases show that a \hyltl query is not satisfied, notably without performing any state-space search. If the LP check is inconclusive, we run our \hyltl engine to perform the state-space exploration using nested depth-first search. Figure~\ref{fig:tapaal-vs-tapaal-LP} shows the ratio plot of TAPAAL \hyltl vs.\ TAPAAL \hyltl\!\!+LP. Most of the additionally answered queries are negative ones that are solved using this LP check.
Even though the LP check is in general often beneficial, the tool allows the user to skip it
and proceed directly to the state-space search if needed.

\begin{figure}[t]
    \centering
    \includegraphics[width=0.8\linewidth]{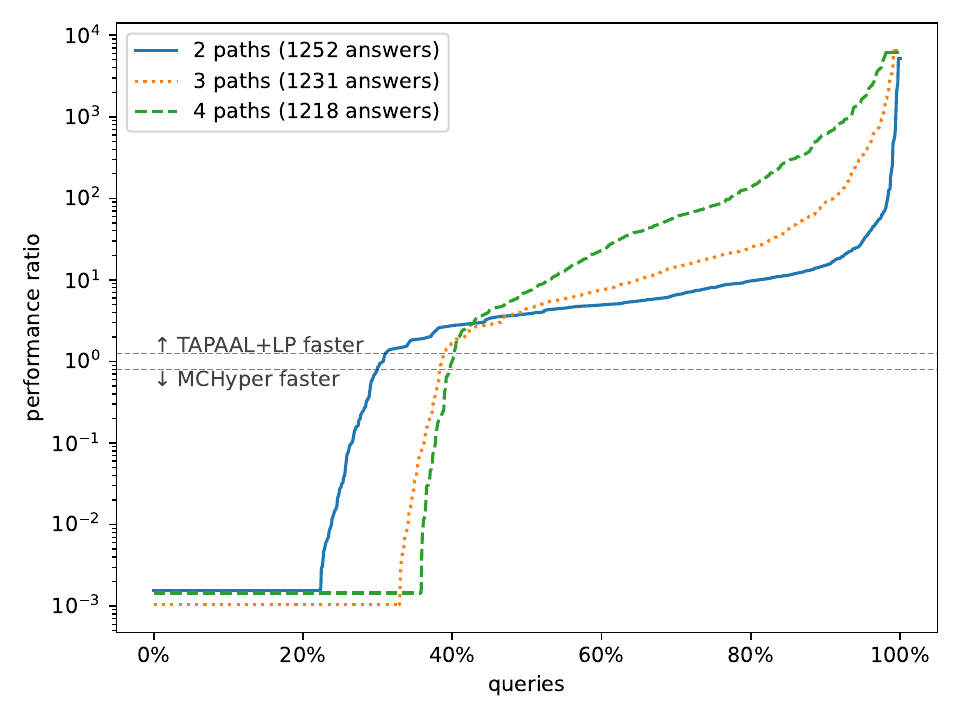}
    \caption{Ratio plot of TAPAAL \hyltl\!\!+LP vs.\ MCHyper}
    \label{fig:tapaal-vs-mchyper}
\end{figure}

Finally, we compare our \hyltl engine with the LP check against MCHyper~\cite{DBLP:conf/cav/CoenenFST19} which
is a state-of-the-art model checker for \hyltl properties. MCHyper operates by encoding the system, described as an AIG circuit, and the formula into a new compact circuit of linear size w.r.t. to the size of the model and the formula~\cite{DBLP:conf/cav/FinkbeinerRS15}. 
While Petri nets can naturally represent nonnegative integers as a number of tokens in places, in MCHyper these numbers have to be encoded into Boolean variables.
To this end, we translated all 260 network topologies in our benchmark into AIG circuits, using a unary encoding for the nodes of the topologies.
The transitions of the system are modeled as a simple state machine, where each transition~$t$ additionally requires a latch, representing the token in the $a(t)$ place, to be enabled.
To allow traces of differing lengths, the circuit cannot leave the target state after entering it (similarly to the Petri net encoding).
The formula is translated into the MCHyper format. For a given problem with parameters $(k, \ell)$, we encode the sum by simply enumerating the $\binom{k}{\ell}$ terms, as the values are sufficiently small.
The comparison of TAPAAL \hyltl\!\!+LP vs.\ MCHyper is provided in Figure~\ref{fig:tapaal-vs-mchyper}. It shows that our tool is faster on about 60\% of all queries, however, MCHyper
solves a significant number of queries where our tool timeouts. This is caused by
the fact that for alternation-free formulae (like in our benchmarks), the encoding of MCHyper allows the property to be verified by a simpler reachability query on the circuit. This enables it to rely on the specialized verification tool ABC~\cite{abc}, which implements state-of-the-art SAT-solvers including PDR (property-directed reachability heuristics)~\cite{pdr}.

\section{Latency Case Study}

As an additional case study (not presented in the conference version of the paper), we consider another variant of routing in computer networks where every link in the network has a certain latency and we must guarantee that for every (loop-free) routing 
from a source to a target we may choose, the difference in the accumulated latency is less than some given latency bound $\ell$. This property is important in computer networking to ensure a high throughput and avoid packet reorderings (which negatively affect, e.g., congestion control algorithms)~\cite{dixit2013impact,mittal2018revisiting}.  The negation of this property can be easily modelled as a \hyltl formula
that asks if there are two loop-free traces that both reach the target node, and at the same time the absolute difference in the latency in these two traces is more than $\ell$.

\begin{figure}[t]
    \centering
\begin{tikzpicture}[font=\scriptsize, xscale=0.25, yscale=0.25, x=1.33pt, y=1.33pt]
\tikzstyle{arc}=[->,>=stealth,thick]
\tikzstyle{transportArc}=[->,>=diamond,thick]
\tikzstyle{inhibArc}=[->,>=o,thick]
\tikzstyle{every place}=[minimum size=6mm,thick]
\tikzstyle{every transition} = [fill=black,minimum width=2mm,minimum height=5mm]
\tikzstyle{every token}=[fill=white,text=black]
\tikzstyle{sharedplace}=[place,minimum size=7.5mm,dashed,thin]
\tikzstyle{sharedtransition}=[transition, fill opacity=0, minimum width=3.5mm, minimum height=6.5mm,dashed]
\tikzstyle{urgenttransition}=[place,fill=white,minimum size=2.0mm,thin]
\tikzstyle{uncontrollabletransition}=[transition,fill=white,draw=black,very thick]
\tikzstyle{globalBox} = [draw,thick,align=left]
\node[place, label={[align=left,label distance=0cm]90:${v_0}$}] at (510,-60) (V0) {};
\node at (510.0,-60.0)[circle,fill,inner sep=1.0pt]{};
\node[place, label={[align=left,label distance=0cm]270:${v_1}$}] at (510,-525) (V1) {};
\node[place, label={[align=left,label distance=0cm]0:${v_2}$}] at (210,-300) (V2) {};
\node[place, label={[align=left,label distance=0cm]0:${v_3}$}] at (885,-300) (V3) {};
\node[place, label={[align=left,label distance=0cm]90:${once(v_1)}$}] at (660,-300) (once_V1) {};
\node at (660.0,-300.0)[circle,fill,inner sep=1.0pt]{};
\node[place, label={[align=left,label distance=0cm]80:${once(v_2)}$}] at (315,-300) (once_V2) {};
\node at (315.0,-300.0)[circle,fill,inner sep=1.0pt]{};
\node[place, label={[align=left,label distance=0cm,xshift=-2mm]20:${once(v_3)}$}] at (735,-300) (once_V3) {};
\node at (735.0,-300.0)[circle,fill,inner sep=1.0pt]{};
\node[place, label={[align=left,label distance=0cm]270:$\mathit{counter}$}] at (15,-480) (counter) {};
\node[transition, label={[align=left,label distance=0cm]120:${t_0}$}] at (555,-300) (T0) {};
\node[transition, label={[align=left,label distance=0cm]120:${t_4}$}] at (660,-375) (T4) {};
\node[transition, label={[align=left,label distance=0cm]120:${t_1}$}] at (315,-135) (T1) {};
\node[transition, label={[align=left,label distance=0cm]350:${t'_4}$}] at (735,-465) (T4_prime) {};
\node[transition, label={[align=left,label distance=0cm]70:${t_2}$}] at (735,-135) (T2) {};
\node[transition, label={[align=left,label distance=0cm]270:${t_3}$}] at (315,-450) (T3) {};
\draw[arc,pos=0.5] (V0) to node[bend right=0,auto,align=left] {} (T0);
\draw[arc,pos=0.5] (V0) to node[bend right=0,auto,align=left] {} (T1);
\draw[arc,pos=0.5] (V0) to node[bend right=0,auto,align=left] {} (T2);
\draw[arc,pos=0.5] (V1) to node[bend right=0,auto,align=left] {} (T4_prime);
\draw[arc,pos=0.5] (V1) to node[bend right=0,auto,align=left] {} (T3);
\draw[arc,pos=0.5] (V3) to node[bend right=0,auto,align=left] {} (T4);
\draw[arc,pos=0.5] (T0) to node[bend right=0,auto,align=left] {} (V1);
\draw[arc,pos=0.5] (T4) to node[bend right=0,auto,align=left] {} (V1);
\draw[arc,pos=0.5] (T1) to node[bend right=0,auto,align=left] {} (V2);
\draw[arc,pos=0.5] (T4_prime) to node[bend right=0,auto,align=left] {} (V3);
\draw[arc,pos=0.5] (T2) to node[bend right=0,auto,align=left] {} (V3);
\draw[arc,pos=0.5] (T3) to node[bend right=0,auto,align=left] {} (V2);
\draw[arc,pos=0.5] (once_V1) to node[bend right=0,auto,align=left] {} (T4);
\draw[arc,pos=0.5] (once_V1) to node[bend right=0,auto,align=left] {} (T0);
\draw[arc,pos=0.5] (once_V2) to node[bend right=0,auto,align=left] {} (T3);
\draw[arc,pos=0.5] (once_V2) to node[bend right=0,auto,align=left] {} (T1);
\draw[arc,pos=0.5] (once_V3) to node[bend right=0,auto,align=left] {} (T2);
\draw[arc,pos=0.5] (once_V3) to node[bend right=0,auto,align=left] {} (T4_prime);
\draw[arc,pos=0.5] (T3) to node[bend right=0,auto,align=left] {} (counter);
\draw[arc,pos=0.1,above] (T2) to node[bend right=0,auto,align=left,above] {$\mathit{3}$ } (counter);
\draw[arc,pos=0.5] (T4_prime) to node[bend right=0,auto,align=left] {} (counter);
\draw[arc,pos=0.2] (T1) to node[bend right=0,auto,align=left,above] {$\mathit{3}$ } (counter);
\draw[arc,pos=0.1] (T4) to node[bend right=0,auto,align=left,above] {$\mathit{4}$ } (counter);
\draw[arc,pos=0.2] (T0) to node[bend right=0,auto,align=left,above] {$\mathit{3}$ } (counter);

\end{tikzpicture}
    \vspace{-10mm}
     $$\exists \pi_1. \exists \pi_2. \F (\pi_1.v_2 = 1\ \wedge\  \pi_2.v_2 = 1\ \wedge\ (\pi_1.counter - \pi_2.counter \ge \ell))$$
    \caption{Latency encoding as a Petri net}
    \label{fig:tapaal-latency}
\end{figure}

\subsection{TAPAAL \hyltl Encoding}
To encode the latency problem in TAPAAL \hyltl, we assume as before a natural way to model packet hops using transition firings. The Petri net model
for our running example from Figure~\ref{fig:network} where the source node
is $v_0$ and the target node is $v_2$ is presented
in Figure~\ref{fig:tapaal-latency}. Additionally, in order to guarantee
loop-freedom, for every place $p$ we add a new place $once(p)$ with a single
token in its initial marking, such that any hop (transition firing)
that places a token to $p$ will also consume a token from $once(p)$.
This guarantees that we never visit the same place twice. To this end, all
transitions that add tokens to the source place $v_0$  
or remove tokens from the target place $v_2$ 
are also removed. In order to measure the latency, we add a new place
$\mathit{counter}$ where every packet hop deposits a number of tokens
that corresponds to the latency of the hop. For example firing $t_2$
deposits $3$ tokens to the place $\mathit{counter}$ to represent
latency $3$ of the hop between the nodes $v_0$ and $v_3$.
The \hyltl formula checking if there are two loop-free traces that
reach $v_2$ with a latency difference larger than or equal to $\ell$
is given under the net in Figure~\ref{fig:tapaal-latency}.

\begin{figure}[t]
    \centering
    \includegraphics[width=0.8\linewidth]{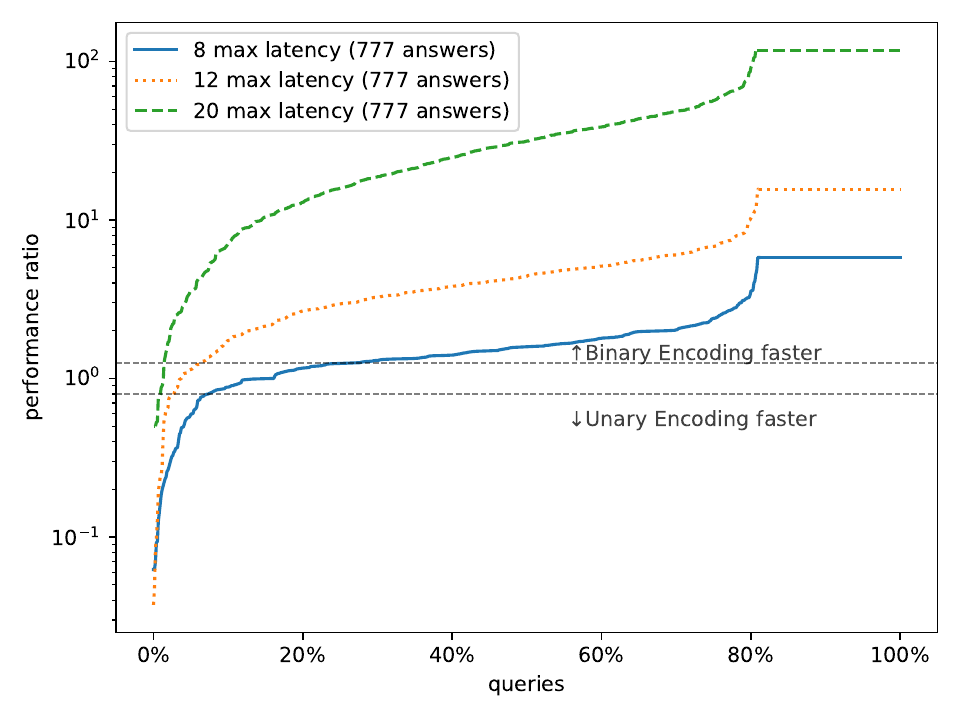}
    \caption{Ratio plot of CPU time of Unary vs.\ Binary Encoding}
    \label{fig:binary-vs-unary}
\end{figure}

\subsection{MCHyper Encoding}
The basic network topology is encoded in the same way as in the previous section.
We use two different encodings of the accumulated latency counter: the counter values are encoded either in binary or in unary.
In the binary encoding we use for efficiency reasons three counters $C_1$, $C_2$ and $C_3$. The first counter $C_1$  uses $\lceil \text{log}_2(\ell)\rceil$ bits and stores the remainder of the accumulated latency divided
by the latency bound $\ell$. The second counter $C_2$ uses $\lceil\text{log}_2(\frac{w_{\text{max}}}{\ell}|E|)\rceil$ bits where $w_{\text{max}}$ is the maximum latency on a link and stores the quotient of the division. For technical reasons, we also have a counter $C_3$;
its value is always equal to the value of $C_2$ plus $1$.

Given two traces $\pi_1$ and $\pi_2$, we formulate the latency query (violation of the latency bound) as: \begin{equation*}
\exists \pi_1.\ \exists \pi_2. \\ \F (\ \pi_1.goal \land \pi_2.goal\land ((\pi_1.C_2 > \pi_2.C_3 )\lor(\pi_1.C_2 = \pi_2.C_3 \land \pi_1.C_1 \geq \pi_2.C_1)))  \ .
\end{equation*} \\
The unary encoding instead does not utilize any counters. Each node is split into multiple nodes based on the weight of the transition. Then, connections are added to the necessary intermediate nodes to accurately represent the latency.
The latency query is then simply \begin{align*}
\exists \pi_1.\ \exists \pi_2. \ \pi_1.goal \land \underbrace{X \cdots X}_{\ell \text{ times}}\neg\pi_2.goal \ .
\end{align*} After the shorter trace reaches the goal, we simply ask that after $\ell$ steps, the other trace is not in the goal state yet.

\subsection{Experimental Setup}
We generated 3891 instances from the topology dataset with random latency values for the links with values between 1 to 5 (in some experiments multiplied by some factor $M$ in order to scale to larger numbers). We use three different values of the latency bound $\ell$, namely 8, 12 and 20, each having 1297 instances.

The experiments are once again executed on an AMD EPYC 7551 processor running at 1996~MHz, with a  900 second timeout and 2GB available memory.

\begin{figure}[t]
\centering
\includegraphics[width=0.8\linewidth]{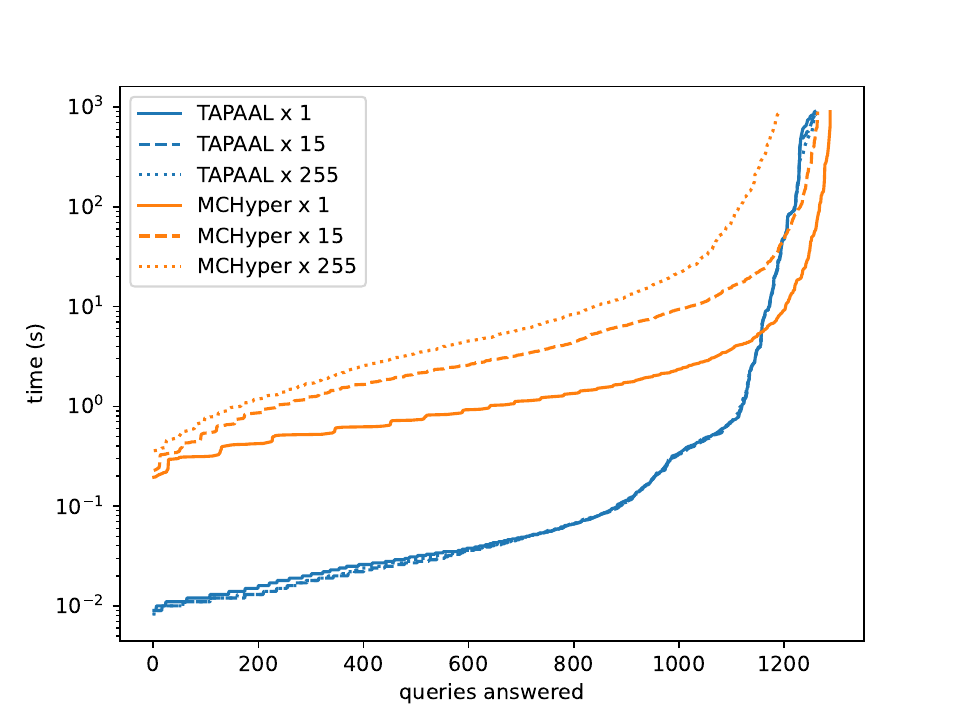}
    \caption{Cactus plot of TAPAAL \hyltl vs MCHyper on multiple scaling factors}
    \label{fig:tapaal-vs-mchyper-scaling}
\end{figure}
\begin{figure}[t]
    \centering
    \includegraphics[width=0.8\linewidth]{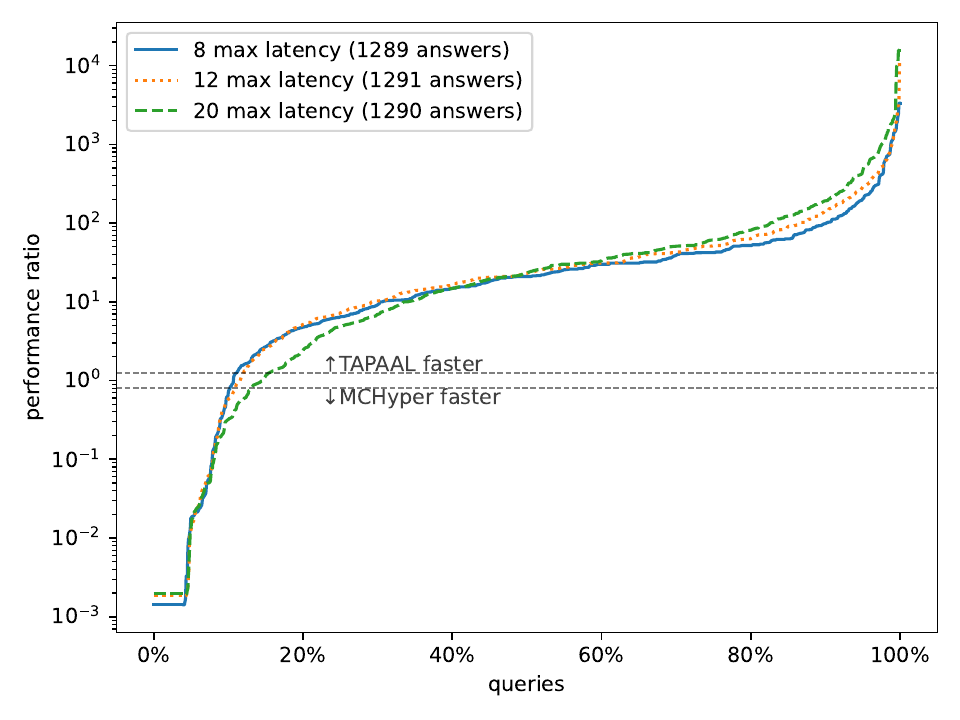}
    \caption{Ratio plot of MCHyper vs.\ TAPAAL \hyltl}
    \label{fig:tapaal-vs-mchyper-latency}
\end{figure}

\subsection{Binary vs.\ Unary Encoding in MCHyper}

Figure~\ref{fig:binary-vs-unary} shows a ratio plot where the CPU time of answering the \hyltl query using the unary encoding divided by the CPU time used by the binary encoding.
The three curves use the three different latency bounds and clearly the binary encoding is faster on more than 80\% of queries and as expected the difference is more pronounced as the latency bound increases as the unary encoding of the problem produces larger transition systems and formulas.
In the remaining experiments, we shall so use the binary encoding in MCHyper.

\subsection{Scaling of Constants and TAPAAL \hyltl vs.\ MCHyper Comparison}

Figure~\ref{fig:tapaal-vs-mchyper-scaling} shows the CPU time (on y-axis)
for all independently sorted instances that where solved by TAPAAL \hyltl and MCHyper.
For each tool, we consider three benchmarks: one with $\ell=8$ and where all latency constants are between 1 to 5 and two more where all constants as well as $\ell$ are multiplied by $15$ respectively $255$. This multiplication creates instances
that are equivalent to the basic one, however, the numbers in the model become larger.

As the TAPAAL \hyltl engine represents numbers as first-class citizens, there is
virtually no difference in the performance as the constant sizes scale up. On the other
hand, MCHyper clearly suffers on the problems with larger constants, despite that it uses the binary encoding. We can also observe that for the smaller constants, MCHyper solves a few more
instances than TAPAAL, however, this is not the case anymore for the benchmark with the large constants.

Finally, Figure~\ref{fig:tapaal-vs-mchyper-latency} depicts the sorted ratios
of MCHyper CPU time divided by TAPAAL \hyltl CPU time on the three benchmarks
(with small constants). On a large majority of cases, TAPAAL \hyltl is faster while
MCHyper solves a few instances on which TAPAAL \hyltl times out. However, in summary,
TAPAAL \hyltl shows even better performance on the latency problems than
on the congestion-free routing problems discussed in Figure~\ref{fig:tapaal-vs-mchyper}.

\section{Conclusion}

We presented the first \hyltl verification engine for Petri nets,
implemented a GUI that allows the user to visually design Petri net
models as well as \hyltl queries and provides debugging feedback 
as the traces discovered by our engine can be simulated in the TAPAAL GUI.
We showed that our \hyltl engine is more efficient than an alternative
self-composition approach and that it is competitive with the state-of-the-art
\hyltl model checker MCHyper. 
In  future work, we plan to transfer
the techniques that enable MCHyper to quickly answer positive \hyltl
queries, in particular property-directed search heuristics, into TAPAAL, in order to further improve its performance.

\paragraph{Acknowledgements.}
Research supported by 
the German Research Foundation (DFG) project, Schwerpunktprogramm SPP 2378: Resilience in Connected Worlds: Mastering Failures, Overload, Attacks, and the Unexpected, project ReNO-2 (1-5003036), 2025-2028, and by
DIREC---Digital Research Centre Denmark.

\bibliographystyle{plain}
\bibliography{bib}

\end{document}